\documentclass{article}
\usepackage{arxiv}
\usepackage[T1]{fontenc} 
\usepackage{biblatex}
\usepackage[utf8]{inputenc}    
\usepackage{hyperref}       
\usepackage{url}            
\usepackage{amsfonts}       
\usepackage{nicefrac}       
\usepackage{microtype}      
\usepackage{graphicx}
\usepackage{doi}
\usepackage{amsmath}
\usepackage{pgfplots}

\title{\emph{Time in a Bottle}: A Psychophysics study of human time perception through aging}

\author{ \href{https://orcid.org/0009-0006-5035-3201}{\includegraphics[scale=0.06]{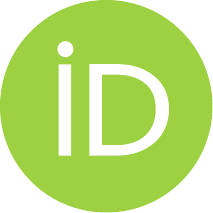}\hspace{1mm}Enric Espel Sanchez}\\
	Department of Physics\\
	University of Helsinki\\
	Helsinki, Kumpula \\
}
\renewcommand{\shorttitle}{\textit{Time In A Bottle}}
\hypersetup{
pdftitle={Time In A Bottle},
pdfsubject={neurobiophysics},
pdfauthor={Enric Espel Sanchez},
pdfkeywords={Neurophysics, Time percpetion, Psychophysics},
}

\pgfplotsset{compat=1.18}
\begin{document}
\maketitle

\begin{abstract}

	Time perception is crucial for a coherent human experience. As life progresses, our perception of the passage of time becomes increasingly non-uniform, often feeling as though it accelerates with age. While various causes for this phenomenon have been theorized, \cite{bejan2019time} a comprehensive mathematical and theoretical framework remains underexplored. This study aims to elucidate the mechanisms behind perceived time dilation by integrating classical and revised psychophysical theorems with a novel mathematical approach. Utilizing Weber's and Fechner's laws as foundational elements, we develop a model that transitions from exponential to logarithmic functions to represent changes in time perception across the human lifespan. Our results indicate that the perception of time shifts significantly around the age of mental maturity, aligning with a proposed inversion point where sensitivity to temporal stimuli decreases, eventually plateauing out at a constant rate. This model not only explains the underlying causes of time perception changes but also provides analytical values to quantify this acceleration. These findings offer valuable insights into the cognitive and neurological processes influencing how we experience time as we go through life.
 
\end{abstract}


\section{Introduction}

The sensation of time speeding up as we age is a phenomenon that concerns everyone on some level, a seemingly universally shared aspect of human existence. Often described as a "quickening" of time with the advent of later years, this dilation seems to abbreviate our experienced lifespan, lending a regrettably poignant brevity to the years that we often describe as flying by. The scientific exploration of this perception not only delves into the roots of human cognition and psychology but also bridges across to our philosophical and existential understanding of experiencing life itself. Historically, the study of time perception has held significant importance in both philosophical inquiries and psychological research. In more recent centuries, this contemplation could be moved into the realm of empirical science, particularly with the advent of psychophysics in the 19th century.\cite{stevens1957psychophysics}

Psychophysics, the branch of study that deals with the relationships between physical stimuli and the resulting perceptions, provided the first systematic approach to studying how we perceive sensations.

\hyperref[sec:Modeling]{See Section 2.}

\subsection{Weber and Fechner}

Ernst Heinrich Weber, a German physician, and one of the early pioneers in psychophysics, was instrumental in laying the foundational concepts of sensory perception. Weber's most notable contribution, now known as Weber's Law, quantifies how the smallest noticeable difference in a stimulus is a constant ratio depending on the initial stimulus. This insight into differential sensitivity was groundbreaking because it established the first quantifiable link between a change in a physical stimulus and our perception of this change.

Weber's Law states that the just noticeable difference ($\Delta I$) in a stimulus is proportional to the magnitude of the original stimulus. Building on Weber's foundational work, Gustav Theodor Fechner, a German experimental psychologist, took these concepts further to develop what is now known as the Weber-Fechner Law. This law posits that the perceived intensity of a stimulus grows logarithmically as the actual intensity grows exponentially. Fechner's work connected measurable external stimuli with the logarithmic scales of human perception, creating a cornerstone for modern psychophysics. Thus, Fechner's Law effectively builds on Weber's Law by suggesting that the perceived intensity (S) grows logarithmically as the actual intensity increases linearly:

\begin{equation}
\Delta I = k \cdot I
\end{equation}

$\Delta I$ is the just noticeable difference,
I is the initial intensity of the stimulus,
k is a constant (Weber's constant).

\begin{equation}
S = k \log \left(\frac{I}{I_0}\right)
\end{equation}

In this case, S is the perceived intensity, I is the actual intensity of the stimulus, I0 is the threshold intensity (the smallest detectable intensity of the stimulus), k is again a constant. Although Weber's theorem does play a part in our perception of time, this has been proven for immediate perception in the small range of milliseconds to seconds. The role of psychophysics in a holistic whole-life time perspective and a mathematical model remain underexplored. \cite{andrew2020time}

\section{Modeling our existence}
\label{sec:Modeling}

\subsection{The Early Years}

The early years of human life hold a profound significance in our perception of existence, marked by a rich tapestry of first experiences and rapid cognitive development. \cite{robyn2003child} Psychologically, these years are perceived as longer and more densely packed with memories, a phenomenon that can be partly explained by the way our brains encode new information.

Novelty plays a critical role in how we perceive time. \cite{wittmann2009inner} \cite{droit2007how}The brain tends to pay more attention to and spends more resources processing new information. Each new experience requires our cognitive frameworks to expand and adapt, processes that are mentally intensive and memorable. Logically then, an exponential model would suggest that the rate at which we encounter new stimuli is initially very high, leading to a perception of time that feels extended.

Utilizing the idea of JND, initially, a year constitutes a large percentage of our memory banks, positing that time perception can only truly be understood (in the long run) via the storage of memories that have filled this previous time. Thus the JND is large compared to the sample size, in short, adding one year is quite noticeable because the overall sample is growing by 100 percent. I.e. An initial exponential increase.

We formulate equation (3):

\begin{equation}
	S(I) = a \cdot e^{bI}
\end{equation}

Where:

S(I) represents the sensation or perceived impact of experiences over time. a and b are constants, with a adjusting the initial sensitivity and b representing the rate of decrease in sensitivity. I is the duration or age in years.

In this formula, b is typically a negative value, indicating that the sensitivity to new experiences decreases exponentially. The mechanisms behind this decrease are primarily twofold. Firstly it mirrors the psychological observations that as we grow older and accumulate more experiences, similar new experiences have less of an impact on us, leading to a faster subjective passage of time.

Secondly and more vitally to this article, the "value" a single year holds in our brain, decreases. I.e. one year represents a smaller and smaller fraction of the whole of our memory bank. Subsequently, keeping with our theory that the mind treats temporal stimuli like physical ones (weight etc.) the JND for time changes and time "speeds up".

\subsection{The Later Years and Speeding Up}

As mentioned, advancing into the later years of our lives, the perception that time is speeding up becomes more pronounced. To prove this and provide a model, we will use Fechner's Law Eq. (2), we will do so to generate a logarithmic perception. As we age, each subsequent year forms a smaller percentage of our total life. Applying and according to Fechner’s Law, this results in a logarithmic decrease, this is desirable for a myriad of reasons:

The JND: Now a single year is smaller than the JND, therefore it takes multiple years for the human mind to accurately gauge that a single year has passed, much like with weight. We posit our minds are processing time perception just like any other stimulus, using memories as the baseline and creating a JND or proprioceptive threshold with every passing year.

Memory Consolidation: Additionally but not as crucially, older adults tend to have fewer new experiences compared to younger individuals, which can affect how memories are formed and recalled. This lack of novelty means fewer distinct memories are created, leading to a sense that less is happening and time is moving faster.

Neurological Factors: Changes in the brain’s neurological structures and functions as we age also play a role. For instance, alterations in dopaminergic pathways, which are crucial for encoding new memories, might lead to a compressed sense of time. \cite{berry2016dopamine}

\subsection{Connecting the Dots: The Inversion Point}

This is the crucial point, unimaginatively dubbed the inversion point, our perception slowly transitions due to each new year approaching the JND. For further calculations, the inversion mean average is taken to be around the age of mental maturity. Up until this threshold (20-30), neural plasticity and connectivity is similar to the early years phase. \cite{steinberg2014age} \cite{arnett2000emerging}

This point marks the transition in the model from one where each additional unit of time adds significantly to perception, to one where each additional unit of time adds progressively less to perception. To model this transition effectively, we create a weighted function that controls how the perception shifts from exponential to logarithmic behavior over time.

The weighted function w(I) will remain crucial for smoothly transitioning between the two distinct phases. This function is designed to vary between 1 and 0 as time progresses. At younger ages, where the exponential model dominates, w(I) is closer to 1, and as time increases, w(I) approaches 0. Mathematically, this is typically represented by a logistic (sigmoid) function:

\begin{equation}
    w(I) = \frac{1}{1 + e^{c(I - \mu)}}
\end{equation}

Here:

\begin{itemize}
    \item $c$ controls the steepness of the transition,
    \item $\mu$ is the midpoint of the transition.
\end{itemize}

\subsection{Defining the Inversion Threshold}

The midpoint \(\mu\) is calculated as the average of two critical thresholds, \(\epsilon\) and \(\delta\), which represent the beginning and end of the transition phase, respectively. Mathematically, \(\mu\) is defined as:

\begin{equation}
    \mu = \frac{\epsilon + \delta}{2}
\end{equation}

The use of \(\mu\) as the average of \(\epsilon\) and \(\delta\) ensures that the logistic weighting function transitions smoothly and symmetrically around the midpoint of these thresholds. This midpoint is chosen because it represents the balanced center between the onset of the transition (\(\epsilon\)) and its conclusion (\(\delta\)).

Simply put, we smooth out the inversion point as an interval.

\subsection{Unified Model for Time Perception}

Combining our three crucial formulations (early years, inversion point and logarithmic interpretation, respectively) into a piece-wise modelling function we obtain: 

\begin{equation}
S(I) = 
\begin{cases} 
a \cdot e^{bI} & \text{if } I \leq \epsilon \\
w(I) & \text{for } \epsilon < I < \delta\\
k \log \left(\frac{I}{I_0}\right) & \text{if } I \geq \delta
\end{cases}
\end{equation}

More precisely, however, we can combine it into a singular sensation function $S(I)$. Which integrates all necessary parts our our model:

\begin{equation}
S(I) = \frac{w(I) \cdot a \cdot e^{bI} + (1 - w(I)) \cdot k \log \left(\frac{I}{I_0}\right)}{w(I) + (1 - w(I))}
\end{equation}

To improve its accuracy further, the denominator in the function, $w(I) + (1 - w(I))$, always simplifies to 1 because $w(I)$ and $1 - w(I)$ are complements. Thus we rewrite the function as follows, giving us our complete mathematical model:

\begin{equation}
S(I) = w(I) \cdot a \cdot e^{bI} + (1 - w(I)) \cdot k \log \left(\frac{I}{I_0}\right)
\end{equation}

or:

\[
S(I) = \left( \frac{1}{1 + e^{-c(I-\mu)}} \right) k \log \left( \frac{I}{I_0} \right) + \left( 1 - \frac{1}{1 + e^{-c(I-\mu)}} \right) a \cdot e^{bI}
\]

where:
\begin{align*}
a & : \text{Initial sensitivity to new experiences}, \\
b & : \text{Rate of decrease in sensitivity over time (exponential component)}, \\
k & : \text{Scale factor for the logarithmic component}, \\
I_0 & : \text{Reference point for the logarithmic component}, \\
c & : \text{Steepness of the transition between exponential and logarithmic phases}, \\
\mu & : \text{Midpoint of the transition}.
\end{align*}

A graphical representation of this function using arbitrary parameters for variables (discussed in the following subsection) can be found below as Fig. 1.

\subsection{Graphical Exploration}

Our graph gives us a good look into our model at a functional level, on the early stages of life every single year (x-axis) corresponds to less than a year in elapsed time (y-axis). This gives the impression that every year is "dragging" on very slowly, their values equalizing at some point during preteen or teen age. This perception is at its highest increase until the inversion point where it slowly transitions to a logarithmic representation. 

Time at this point seems to have significantly sped up, 8 fold what was given at early stages, eventually reaching an asymptotical plateau. Although when this would be reached seems to outlast current human lifespan. Perhaps this plateau is a mathematical representation of the result of how we perceive time, eventually reaching a JND so transformed it is biologically impossible to alter it further.

For this exact graph, the following values for constants were used: a = 0.1, b = -0.1, k = 2, Izero = 1, c = 0.2, $\mu$ = 20.

\begin{figure}[ht]
    \centering
    \begin{tikzpicture}
        \begin{axis}[
            width=12cm,
            height=8cm,
            xlabel={Age ($I$)},
            ylabel={Perceived Sensation ($S(I)$)},
            title={Model of Perception of Time Throughout the Human Lifespan},
            grid=major,
            legend pos=north west,
            samples=400,
            domain=0.1:100,
            ymin=0,
        ]

        \pgfmathsetmacro{\a}{0.1}
        \pgfmathsetmacro{\b}{-0.1}
        \pgfmathsetmacro{\k}{2}
        \pgfmathsetmacro{\Izero}{1}
        \pgfmathsetmacro{\c}{0.2}
        \pgfmathsetmacro{\mu}{20}

        \addplot[
            thick,
            blue
        ] (
            x, 
            {
                (1 / (1 + exp(-\c * (x - \mu)))) * \k * ln(x / \Izero) + 
                (1 - (1 / (1 + exp(-\c * (x - \mu))))) * \a * exp(\b * x)
            }
        );
        \addlegendentry{Perception of Time ($S(I)$)}

        \end{axis}
    \end{tikzpicture}
    \caption{Model of Perception of Time Throughout the Human Lifespan}
    \label{fig:time_perception}
\end{figure}
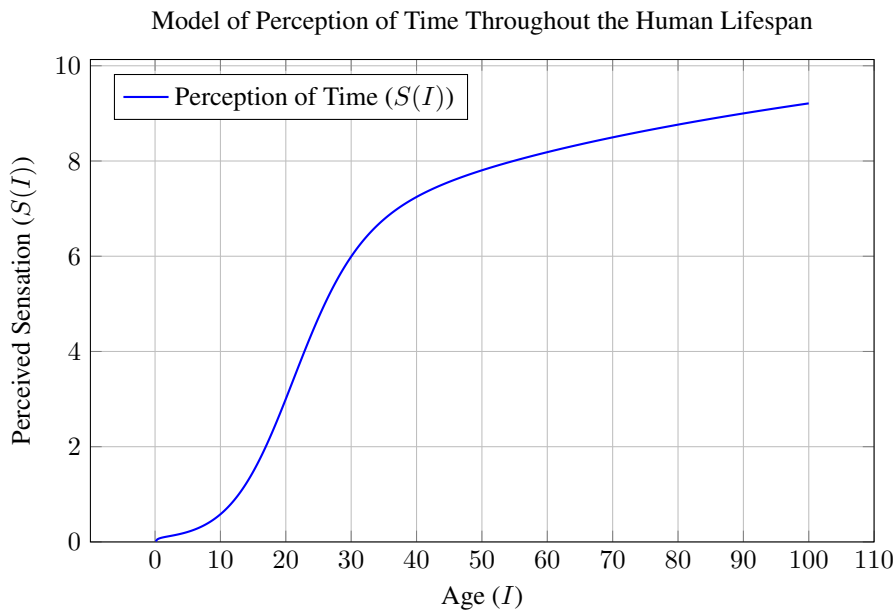

\subsection{The Additive Approach, a Flexible Model}

Further enhancing the model would involve a multi-faceted approach that integrates deeper neurological insights, further advanced mathematical techniques, and a broader consideration of cognitive and behavioral factors. By incorporating elements such as dopaminergic pathways, synaptic pruning, neuroplasticity, cognitive load, and emotional impacts, the model could maybe more accurately reflect the complex interplay of factors influencing how humans perceive time.

Leveraging stochastic modeling, and machine learning could further refine the model, allowing it to capture dynamic and individualized aspects of time perception. However, it is crucial to balance complexity with interpretability, ensuring that the model remains both scientifically robust and comprehensible. Cluttering the model with a myriad of mathematical terms, whilst helping real-world exactitude, could prove to be detrimental to the usefulness of the model and theorem themselves.

Ultimately, continuous validation with empirical data and iterative refinement based on new scientific discoveries will be essential in further advancing this into a comprehensive and accurate model of human time perception through the lifespan.

\section{Conclusion}

In conclusion, what we have developed is a mathematical approach to a common seemingly subjective aspect of human consciousness. This mathematical model, however, should not be taken as an absolute but merely as a theoretical tool to introduce psychophysics and more robust mathematics to temporal problems. Human consciousness, and time perception by extension, are a largely unknown phenomenon, and much of the work done is speculative.

Our graphical representation, however, affirms the theoretical assertions, showcasing a clear demarcation between the two perceptual phases. Interestingly, the asymptotic plateau observed suggests a biological and psychological ceiling in time perception, beyond which additional aging yields minimal perceptual changes.

\printbibliography

\end{document}